\documentclass[aps,showpacs,preprintnumbers,nofootinbib]{revtex4}
\usepackage{amssymb}
\usepackage{amsfonts}
\usepackage{amsmath}
\usepackage{graphicx}
\usepackage{dcolumn}
\usepackage{bm}
\usepackage[latin1]{inputenc}
\usepackage[dvips]{hyperref}
\usepackage{graphicx}

\def\be{\begin{equation}}
\def\ee{\end{equation}}
\def\bea{\begin{eqnarray}}
\def\eea{\end{eqnarray}}

\begin{document}

\title{Scalar field perturbations of a Lifshitz black hole in conformal gravity in three dimensions%
}
\author{Marcela Catal\'an}
\email{marceicha@gmail.com}
\affiliation{Departamento de Ciencias F\'isicas, Facultad de Ingenier\'ia y Ciencias,
Universidad de La Frontera, Avenida Francisco Salazar 01145, Casilla 54-D,
Temuco, Chile.}

\author{Yerko V\'asquez}
\email{yvasquez@userena.cl}
\affiliation{Departamento de F\'isica, Facultad de Ciencias, Universidad de La Serena,\\
Avenida Cisternas 1200, La Serena, Chile.}
\date{\today }

\begin{abstract}
We study the reflection and transmission coefficients and the absorption cross section for scalar fields in the background of a  Lifshitz black hole in three-dimensional conformal gravity with $z=0$, and we show that the absorption cross section vanishes at the low and high frequency limit. Also, we determine the quasinormal modes of scalar perturbations
and then we study the stability of these black holes under scalar field
perturbations.

\end{abstract}

\maketitle


\section{Introduction}

The three-dimensional models of gravity have attracted a remarkable interest in recent years. Apart from the BTZ black hole \cite%
{Banados:1992wn}, which is a solution to the Einstein equations with a
negative cosmological constant, much attention has been paid to
topologically massive gravity (TMG), which generalizes
three-dimensional general relativity (GR) by adding a Chern-Simons gravitational term to the Einstein-Hilbert action \cite{Deser:1981wh}. In this model, the
propagating degree of freedom is a massive graviton. TMG also admits the BTZ and other black holes as exact solutions. The renewed interest in TMG
results from the possibility of constructing a chiral theory of gravity at a
special point of the space of parameters, \cite{Deser:1982vy}.

Another three-dimensional massive gravity theory that has attracted attention in recent years is known as new massive gravity
(NMG), where the action is the standard Einstein-Hilbert term plus a specific
combination of scalar curvature square term and a Ricci tensor square term \cite%
{Bergshoeff:2009hq, Bergshoeff:2009tb,
Andringa:2009yc}, and at the linearized
level it is equivalent to the Fierz-Pauli action for a massive spin-2 field \cite%
{Bergshoeff:2009hq}. NMG admits interesting solutions, see for instance \cite{Clement:2009gq, AyonBeato:2009yq, Clement:2009ka, AyonBeato:2009nh, Oliva:2009ip, Correa:2014ika}; for further aspects of NMG see \cite{Nakasone:2009bn, Bergshoeff:2009aq,
Oda:2009ys, Deser:2009hb}. TMG and NMG share common features, however there are different
aspects: one of these is the existence in NMG of black holes known as new type black holes, which are also solutions of conformal gravity in three dimensions \cite{Oliva:2009hz}.

Lifshitz spacetimes have received considerable attention from the point of view of
condensed matter, i.e. the search for gravity duals of
Lifshitz fixed points due to the AdS/CFT correspondence \cite{Maldacena:1997re} for condensed matter
physics and quantum chromodynamics \cite{Kachru:2008yh, Hartnoll:2009ns}. There are many invariant scale theories of
interest when studying such critical points, such theories exhibit the
anisotropic scale invariance $t\rightarrow \lambda ^{z}t$, $x\rightarrow \lambda x$, where $z$ is the relative scale dimension of time and space,
and they are of particular interest in studies of critical exponent theory
and phase transitions. Lifshitz spacetimes are described by the metrics 
\begin{equation}
ds^2=- \frac{r^{2z}}{l^{2z}}dt^2+\frac{l^2}{r^2}dr^2+\frac{r^2}{l^2} d\vec{x}%
^2~,  \label{lif1}
\end{equation}
where $\vec{x}$ represents a $d-2$ dimensional spatial vector, with $d$ representing the
spacetime dimension while $l$ denotes the length scale in the geometry. For $z=1$ this spacetime is the usual anti-de Sitter metric in Poincar\'e coordinates.

The metrics of the Lifshitz
black holes asymptotically have the form (\ref{lif1}). Several Lifshitz black holes have been reported \cite%
{AyonBeato:2009nh, Cai:2009ac, AyonBeato:2010tm, Dehghani:2010kd, Mann:2009yx, Balasubramanian:2009rx, Bertoldi:2009vn, Bravo-Gaete:2013dca, Correa:2014ika, Alvarez:2014pra}.

In this work, we will
consider a matter distribution outside the horizon of a Lifshitz black
hole in three-dimensional conformal gravity with $z=0$. It is worth mentioning that for $z=0$ the
anisotropic scale invariance corresponds to
space-like scale invariance with no transformation of time. The matter is
parameterized by a scalar field, which we will perturb by assuming that
there is no back reaction on the metric. We then obtain the reflection and transmission coefficients, the absorption cross section and the
quasinormal frequencies (QNFs) for scalar
fields, and study their stability under scalar perturbations.

Conformal gravity is a four-derivative theory and it is perturbatively renormalizable \cite{Stelle:1976gc, Stelle:1977ry}. Furthermore, it
contains ghostlike modes, in the form of massive spin-2 excitations.
However, it has been shown that by using the method of Dirac constraints \cite%
{Dirac} to quantize a prototype second plus fourth order theory, viz. the
Pais-Uhlenbeck fourth order oscillator model \cite{Pais:1950za}, it becomes apparent that the limit in which the second order piece is
switched off is a highly singular one in which the would-be ghost states
move off the mass shell \cite{Mannheim:2006rd}. In three spacetime dimensions the equations of motion contain third derivatives of the metric. AdS and Lifshitz black holes with $z=0$ and  $z=4$ in four-dimensional conformal gravity have been studied in \cite{Lu:2012xu}, and solutions to conformal gravity in three dimensions have been studied in \cite{Oliva:2009hz, Guralnik:2003we, Grumiller:2003ad}.

Several studies have contributed to the scattering and absorption properties of waves in the spacetime of black holes. As the geometry of the spacetime surrounding a black hole is non-trivial, the Hawking radiation emitted at the event horizon may be modified by this geometry, so that when an observer located very far away from the black hole measures the spectrum, this will no longer be that of a black body \cite{Maldacena:1996ix}. The factors that modify the spectrum emitted by a black hole are known as greybody factors and can be obtained through classical scattering; their study therefore allows the semiclassical gravity dictionary to be increased, and also offers insight into the quantum nature of black holes and thus of quantum gravity; for an excellent review of this topic see \cite{Harmark:2007jy}. Also, see for instance \cite{Moderski:2008nq, Gibbons:2008gg}, for decay of Dirac fields in higher dimensional black holes.

On the other hand, the study of the QNFs gives information about the stability of black holes
under matter fields that evolve perturbatively in their exterior region,
without backreacting on the metric \cite{Regge:1957td, Zerilli:1971wd,
Zerilli:1970se, Kokkotas:1999bd, Nollert:1999ji, Konoplya:2011qq}.
Furthermore, according to the AdS/CFT correspondence the QNFs determine how fast a thermal
state in the boundary theory will reach thermal equilibrium \cite{Birmingham:2001pj}. Also, in \cite{Corda:2012tz, Corda:2012dw} the authors discuss a connection between Hawking radiation and quasinormal modes (QNMs), which can naturally be interpreted in terms of quantum levels. In three dimensions, the QNMs of BTZ black holes have been studied in \cite{Chan:1996yk, Cardoso:2001hn, Birmingham:2001pj}; scalar and fermionic QNMs in the background of new type black holes in NMG were studied
in \cite{Kwon:2011ey} and \cite{Gonzalez:2014voa}, respectively. Studies of QNMs in Lifshitz black holes can be found in \cite{CuadrosMelgar:2011up, Gonzalez:2012de, Gonzalez:2012xc, Myung:2012cb, Becar:2012bj,Giacomini:2012hg, Catalan:2014ama, Catalan:2013eza, Lopez-Ortega:2014daa, Lopez-Ortega:2014oha}. On the other hand, the absoption cross section for Lifshitz black holes were examined in \cite%
{Moon:2012dy, Gonzalez:2012xc, Lepe:2012zf}, and particle motion in these geometries in 
\cite{Olivares:2013zta, Olivares:2013uha, Villanueva:2013gra}.
Fermions on Lifshitz Background have been studied in \cite{Alishahiha:2012nm}.

The paper is organized as follows. In Sec. II we find a Lifshitz black hole with $z=0$ in three-dimensional conformal gravity. In Sec. III we calculate the reflection and transmission coefficients, the absorption cross section and the quasinormal modes of scalar perturbations for the three-dimensional Lifshitz
black hole with $z=0$. We conclude with Final Remarks in Sec. IV.

\section{Lifshitz black hole in three-dimensional conformal gravity with $z=0$}

In this work we will consider a matter distribution described by a scalar
field outside the event horizon of an asymptotically Lifshitz black hole in
three-dimensional conformal gravity with $z=0$. In
three dimensions, the field equations of conformal gravity are given by the vanishing of the Cotton tensor:
\begin{equation} \label{fieldeq}
C^{\alpha}_{\,\,\beta}=\epsilon ^{\rho \sigma \alpha} \nabla_{\rho} \left( R_{\sigma \beta}-\frac{1}{4}g_{\sigma \beta} R\right)=0~,
\end{equation}
where $R$ is the Ricci scalar. The Cotton tensor is a traceless tensor that vanishes if and only if the metric is locally conformally flat. Solutions to this theory have been studied in \cite{Guralnik:2003we, Grumiller:2003ad, Oliva:2009hz}.

In order to find a Lifshitz black hole solution in three-dimensional conformal gravity, we consider the following ansatz for the metric

\begin{equation}
ds^{2} =-r^{2z} f(r )dt^{2}+\frac{dr^{2}}{r^{2}f ( r )}+r^{2}d\theta ^{2}~,
\end{equation}

and from the field equations (\ref{fieldeq}) we obtain

\begin{equation} \label{differential}
4z\left(z-1  \right)f(r)+\left( -3+7z+2z^2 \right)rf^{\prime}(r) +3\left( 1+z \right) r^2f^{\prime \prime}(r) +r^3 f^{\prime \prime \prime} (r)=0~.
\end{equation}

For $f(r)=1$ (Lifshitz spacetime), the above equation yields the solutions $z=0$ and $z=1$. However, as we mentioned, $z=1$ corresponds to the usual anti-de Sitter metric in Poincar\'e coordinates.

Now, doing $z=0$ in the field equation (\ref{differential}), the following Lifshitz black hole solution is found

\begin{eqnarray} \label{solution}
ds^{2} & = & - f(r )dt^{2}+\frac{dr^{2}}{r^{2}f ( r )}+r^{2}d\theta ^{2}~, \notag \\
f ( r ) & = & 1-\frac{M}{r^{2}}~.
\end{eqnarray}

The requirement $r_{+}=\sqrt{M}>0$ implies that $M>0$. The Kretschmann
scalar is given by 
\begin{equation}
R_{\mu \nu \rho \sigma }R^{\mu \nu \rho \sigma }=4+\frac{20M^{2}}{r^{4}},
\end{equation}%
therefore, there is a curvature singularity at $r=0$ for $M\neq 0$. Other asymptotically Lifshitz black hole solutions in four-dimensional and six-dimensional conformal gravity are found in \cite{Lu:2012xu} and \cite{Lu:2013hx}, respectively.

Metric (\ref{solution}) is conformally related to an asymptotically de Sitter black hole. This can be seen by defining $d\bar{s}^2=\Omega^2ds^2$, where the conformal factor is given by

\begin{equation}
\Omega=\frac{2r}{1+2Nr^2}~,
\end{equation}
and $N$ is a constant. The new radial coordinate $\bar{r}$ is defined by $\bar{r}=\Omega r$, and

\begin{equation}
d\bar{s}^{2} =-\bar{f}(\bar{r} )dt^{2}+\frac{d\bar{r}^{2}}{\bar{f} ( \bar{r} )}+\bar{r}^{2}d\theta ^{2}~,
\end{equation}

where

\begin{equation} \label{conf}
\bar{f}(\bar{r})=-2N\left( 1+2MN\right) \bar{r}^2+2\left( 1+4MN \right) \bar{r}-4M ~,
\end{equation}
this metric corresponds to an asymptotically de Sitter black hole. Metrics of the form (\ref{conf})  were studied in \cite{Oliva:2009hz}.
Notice that the conformal factor is not singular on the horizon. However, it is singular at the asymptotic region $r\rightarrow \infty$, therefore, the asymptotic regions of the two conformally related metrics are very different .

In the next section, we will determine the reflection coefficient, the transmission coefficient and the absorption cross section of the Lifshitz black hole metric (\ref{solution}) found in this section. Then, we will compute the QNFs, which coincide with the poles of the transmission coefficient and we will study the linear stability of these black holes under scalar field perturbations.

\section{Reflection coefficient, transmission coefficient, absorption
cross section and quasinormal modes of $z=0$ Lifshitz black hole}

The scalar perturbations in the background of an
asymptotically Lifshitz black hole in three-dimensional conformal gravity with dynamical
exponent $z=0$ are given by the Klein-Gordon equation of the scalar field
solution with suitable boundary conditions at the event horizon and at the asymptotic infinity.

In section $A$ we will consider a scalar field minimally coupled to gravity. Then, in section $B$, we will study a scalar field non-minimally coupled.

\subsection{Scalar field minimally coupled to gravity}

\begin{figure}[h]
\begin{center}
\includegraphics[width=0.7\textwidth]{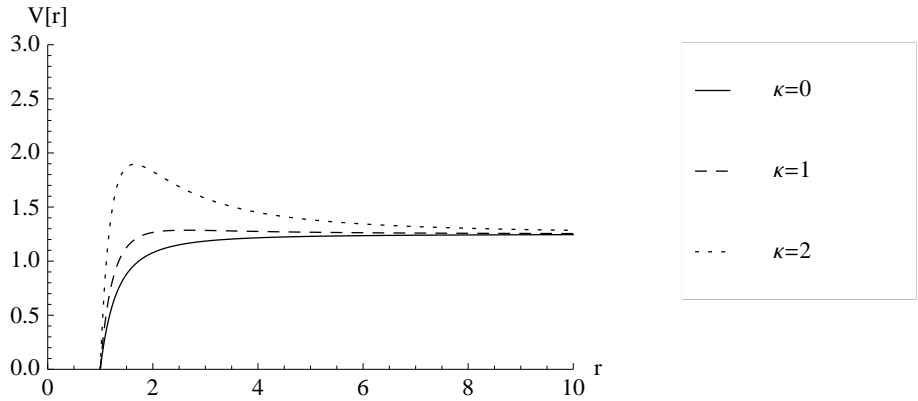}
\end{center}
\caption{The effective potential as a function of $r$, for $M=1$, $m=1$  and $%
\protect\kappa =0, 1, 2$.}
\label{potencial}
\end{figure}

The Klein-Gordon equation in curved spacetime is given by
\begin{equation}
\frac{1}{\sqrt{-g}}\partial _{\mu }\left( \sqrt{-g}g^{\mu \nu }\partial
_{\nu }\right) \phi =m^{2}\phi ~,  \label{KG}
\end{equation}%
where $m$ is the mass of the scalar field $\phi $, which is minimally
coupled to curvature. By means of the following ansatz 
\begin{equation}
\phi =e^{-i\omega t}e^{i\kappa \theta }R(r)~,
\end{equation}%
the Klein-Gordon equation reduces to the following differential equation for the radial function $R(r)$
\begin{equation}
r^{2}\left( r^{2}-M\right) \frac{d^{2}R\left(
r\right)}{dr^{2}}+2r^{3}\frac{dR\left(
r\right)}{dr}+\left( 
\frac{\omega ^{2}r^{4}}{r^{2}-M}-\kappa ^{2}-m^{2}r^{2}\right) R\left(
r\right) =0~.  \label{first}
\end{equation}%
Now, by considering $R(r)=r^{-1/2}G(r)$ and by introducing the tortoise
coordinate $x$, given by $dx=\frac{dr}{rf(r)}$, the latter
equation can be rewritten as a one-dimensional Schr\"odinger equation 
\begin{equation}
\left[ \partial _{x}^{2}+\omega ^{2}-V_{eff}(r)\right] G(x)=0~,
\end{equation}%
where the effective potential is given by 
\begin{equation}
V_{eff}(r)=f(r)-\frac{3}{4}f(r)^{2}+\frac{\kappa ^{2}}{r^{2}}f\left(
r\right) +m^{2}f\left( r\right)~.
\end{equation}%
In Fig. (\ref{potencial}) we plot the effective potential for $M=1$, $m=1$ and different values of the angular momentum $\kappa =0, 1, 2$. Note that, when $r\rightarrow \infty $ the effective
potential goes to $1/4+m^{2}$. 

Performing the change of variable $z=1-\frac{M}{r^{2}}$, the radial equation
(\ref{first}), can be written as 
\begin{equation}
z\left( 1-z\right) \partial _{z}^{2}R(z)+\left( 1-\frac{3}{2}z\right)
\partial _{z}R(z)+\frac{1}{4}\left[ \frac{\omega ^{2}}{z(1-z)}-\frac{\kappa
^{2}}{M}-\frac{m^{2}}{1-z}\right] R(z)=0~.  \label{equationc1}
\end{equation}%
Using the decomposition $R(z)=z^{\alpha }(1-z)^{\beta }F(z)$, with 
\begin{equation}
\alpha _{\pm }=\pm \frac{i\omega }{2}~,
\end{equation}%
\begin{equation}
\ \beta _{\pm }=\frac{1}{4}\left( 1\pm \sqrt{1+4(m^{2}-\omega ^{2})}\right)
~,  \label{omega2}
\end{equation}%
allows us to write (\ref{equationc1}) as a hypergeometric equation for $F(z)$
\begin{equation}
z(1-z)\partial _{z}^{2}F(z)+\left[ c-(1+a+b)z\right] \partial _{z}F(z)-abF(z)=0~,  \label{hypergeometric}
\end{equation}%
where the coefficients are given by 
\begin{equation}
\ a=\frac{1}{4}+\alpha +\beta \pm \frac{1}{4}\sqrt{1-\frac{4\kappa ^{2}}{M}}%
~,  \label{a}
\end{equation}%
\begin{equation}
b=\frac{1}{4}+\alpha +\beta \mp \frac{1}{4}\sqrt{1-\frac{4\kappa ^{2}}{M}}~,
\end{equation}%
\begin{equation}
c=1+2\alpha ~.
\end{equation}%
The general solution of the hypergeometric equation~(\ref{hypergeometric})
is 
\begin{equation}
F\left( z\right) =C_{1}{_{2}}F{_{1}}(a,b,c;z)+C_{2}z^{1-c}{_{2}}F{_{1}}%
(a-c+1,b-c+1,2-c;z)~,
\end{equation}%
and it has three regular singular points at $z=0$, $z=1$, and $z=\infty $. ${%
_{2}}F{_{1}}(a,b,c;z)$ is a hypergeometric function and $C_{1}$ and $C_{2}$
are integration constants. Thus, the solution for the radial function $R(z)$ is 
\begin{equation}
\ R(z)=C_{1}z^{\alpha }(1-z)^{\beta }{_{2}}F{_{1}}(a,b,c;z)+C_{2}z^{-\alpha
}(1-z)^{\beta }{_{2}}F{_{1}}(a-c+1,b-c+1,2-c;z)~.  \label{RV}
\end{equation}%
So, in the vicinity of the horizon, $z=0$ and using the property $%
F(a,b,c,0)=1$, the function $R(z)$ behaves as 
\begin{equation}
R(z)=C_{1}e^{\alpha \ln z}+C_{2}e^{-\alpha \ln z}~,  \label{Rhorizon}
\end{equation}%
and the scalar field $\phi $, for $\alpha =\alpha _{-}$, can be written as
follows 
\begin{equation}
\phi \sim C_{1}e^{-i\omega (t+\ln z)}+C_{2}e^{-i\omega (t-\ln z)}~,
\end{equation}%
where the first term represents an ingoing wave and the second one an
outgoing wave in the black hole. So, by imposing that only ingoing waves
exist at the horizon, this fixes $C_{2}=0$. The radial solution then becomes 
\begin{equation}
R(z)=C_{1}e^{\alpha \ln z}(1-z)^{\beta }{_{2}}F{_{1}}(a,b,c;z)=C_{1}e^{-i%
\omega \ln z}(1-z)^{\beta }{_{2}}F{_{1}}(a,b,c;z)~.  \label{horizonsolution}
\end{equation}

The reflection and transmission coefficients depend on the
behavior of the radial function, both at the horizon and at the asymptotic
infinity, and they are defined by 
\begin{equation}
\mathcal{R}=\left\vert \frac{\mathcal{F}_{\mbox{\tiny asymp}}^{\mbox{\tiny
out}}}{\mathcal{F}_{\mbox{\tiny asymp}}^{\mbox{\tiny in}}}\right\vert
;\;\mathcal{T}=\left\vert \frac{\mathcal{F}_{\mbox{\tiny
hor}}^{\mbox{\tiny in}}}{\mathcal{F}_{\mbox{\tiny asymp}}^{\mbox{\tiny
in}}}\right\vert ~,  \label{reflectiond}
\end{equation}%
where $\mathcal{F}$ is the flux, and is given by 
\begin{equation}
\mathcal{F}=\frac{1}{2i}\sqrt{-g}~g^{rr}\left( R \left(r \right)^{\ast }\frac{dR \left(r \right)}{dr}-R \left(r \right)\frac{%
dR \left(r \right)^{\ast }}{dr}\right)~,  \label{flux1}
\end{equation}
where $
\sqrt{-g}=1$. The behavior at the horizon is given by~(\ref{Rhorizon}), and
using~(\ref{flux1}), we get the flux at the horizon  
\begin{equation}
\mathcal{F}_{hor}^{in}=-\omega \sqrt{M}|C_{1}|^{2}~.
\end{equation}%
Now, in order to obtain the asymptotic behavior of $R \left(z \right) $, we take the limit $z\rightarrow 1 $ in (\ref{equationc1}). Thus, we
obtain the following solution

\begin{equation}
R\left( z\right) =B_{1}\left( 1-z\right) ^{\beta _{+}}+B_{2}\left(
1-z\right) ^{\beta _{-}}~.
\end{equation}%
Thus, the flux~(\ref{flux1}) at the asymptotic region is given by 
\begin{equation}
\ \mathcal{F}_{asymp}=-\sqrt{M}\sqrt{\omega ^{2}-m^{2}-\frac{1}{4}}\left(
\left\vert B_{1}\right\vert ^{2}-\left\vert B_{2}\right\vert ^{2}\right) ~,
\label{fluxdinfinity}
\end{equation}%
where $\omega ^{2}\geq m^{2}+\frac{1}{4}$. On the other hand, by replacing
Kummer's formula \cite{M. Abramowitz}, in (\ref{horizonsolution}), 
\begin{eqnarray}
{_{2}F_{1}}\left( a,b,c;z\right)  &=&\frac{\Gamma \left( c\right) \Gamma
\left( c-a-b\right) }{\Gamma \left( c-a\right) \Gamma \left( c-b\right) }{%
_{2}F_{1}}\left( a,b,a+b-c;1-z\right) +  \notag  \label{KE} \\
&&\left( 1-z\right) ^{c-a-b}\frac{\Gamma \left( c\right) \Gamma \left(
a+b-c\right) }{\Gamma \left( a\right) \Gamma \left( b\right) }{_{2}F_{1}}%
\left( c-a,c-b,c-a-b+1;1-z\right) ~,
\end{eqnarray}%
and by using Eq. (\ref{flux1}) we obtain the flux 
\begin{equation}
\ \mathcal{F}_{asymp}=-\sqrt{M}\sqrt{\omega ^{2}-m^{2}-\frac{1}{4}}\left(
\left\vert A_{1}\right\vert ^{2}-\left\vert A_{2}\right\vert ^{2}\right)~,
\end{equation}%
where, 
\begin{eqnarray}
A_{1} &=&C_{1}\frac{\Gamma \left( c\right) \Gamma \left( a+b-c\right) }{%
\Gamma \left( a\right) \Gamma \left( b\right) }=B_{1}~,  \notag \\
A_{2} &=&C_{1}\frac{\Gamma \left( c\right) \Gamma \left( c-a-b\right) }{%
\Gamma \left( c-a\right) \Gamma \left( c-b\right) }=B_{2}~.
\end{eqnarray}%
Therefore, the reflection and transmission coefficients are given by 
\begin{equation}
\mathcal{R}=\frac{|A_{1}|^{2}}{|A_{2}|^{2}}~,  \label{coef12}
\end{equation}%
\begin{equation}
\mathcal{T}=\frac{\omega |C_{1}|^{2}}{\sqrt{\omega ^{2}-m^{2}-\frac{1}{4}}|A_{2}|^{2}}%
~,  \label{coef22}
\end{equation}%
and the absorption cross section $\sigma _{abs}$, becomes 
\begin{equation}
\ \sigma _{abs}= \frac{\mathcal{T}}{\omega}= \frac{|C_{1}|^{2}}{\sqrt{\omega ^{2}-m^{2}-\frac{1}{4}}%
|A_{2}|^{2}}~.  \label{absorptioncrosssection2}
\end{equation}%

Now, we will study the behavior of the reflection and transmission coefficients and the absorption cross section at the low and high frecuency limits.

At the low frequency limit, $\omega^2 \approx m^2+1/4$, the coefficients $A_{1}$ and $A_{2}$ are given aproximately by

\begin{eqnarray}
A_{1}  & \approx & C_{1} \frac{\Gamma \left(1-i\omega \right) \Gamma \left(i\sqrt{\omega^2-m^2-\frac{1}{4}} \right)}{\Gamma \left(\frac{1}{2}-\frac{i\omega}{2}+\frac{1}{4}\sqrt{1-\frac{4\kappa^2}{M}} \right)\Gamma \left(\frac{1}{2}-\frac{i\omega}{2}-\frac{1}{4}\sqrt{1-\frac{4\kappa^2}{M}} \right)}~,  \notag \\
A_{2}  & \approx & C_{1} \frac{\Gamma \left(1-i\omega \right) \Gamma\left(-i\sqrt{\omega^2-m^2-\frac{1}{4}}\right)}{\Gamma \left(\frac{1}{2}-\frac{i\omega}{2}-\frac{1}{4}\sqrt{1-\frac{4\kappa^2}{M}} \right)\Gamma \left(\frac{1}{2}-\frac{i\omega}{2}+\frac{1}{4}\sqrt{1-\frac{4\kappa^2}{M}} \right)}~.
\end{eqnarray}

Using the expressions above, the following low frequency limit is obtained for the reflection coefficient

\begin{equation}
\mathcal{R} \approx \left|\frac{\Gamma \left(i\sqrt{\omega^2-m^2-\frac{1}{4}} \right)}{\Gamma \left(-i\sqrt{\omega^2-m^2-\frac{1}{4}} \right)}\right|^2 =1~,
\end{equation}

where, in the last equality we used the fact that $\Gamma(z)^*=\Gamma(z^*)$. For
the transmission coefficient and the absorption cross section, we find

\begin{eqnarray}
\mathcal{T}  & \propto & \frac{1}{\sqrt{\omega^2-m^2-\frac{1}{4}}\left|  \Gamma \left(-i\sqrt{\omega^2-m^2-\frac{1}{4}} \right) \right|^2}=\frac{i}{\Gamma \left(-i\sqrt{\omega^2-m^2-\frac{1}{4}} +1\right)\Gamma \left(i\sqrt{\omega^2-m^2-\frac{1}{4}} \right)} \rightarrow 0~, \notag \\
\sigma  & = & \frac{\mathcal{T}}{\omega} \propto \frac{1}{\Gamma \left(i\sqrt{\omega^2-m^2-\frac{1}{4}} \right)}\rightarrow 0~,
\end{eqnarray}

where we have used the properties $z\Gamma(z)=\Gamma(z+1)$, $\Gamma(z)^*=\Gamma(z^*)$ and $\Gamma \left(i\sqrt{\omega^2-m^2-\frac{1}{4}} \right) \rightarrow \infty$.

At the high frequency limit $\omega^2>>m^2+1/4$, the coefficients $A_{1}$ and $A_{2}$ are given by

\begin{eqnarray} \label{AA}
A_{1}  & \approx & C_{1} \frac{\Gamma \left(1-i\omega \right) \Gamma \left(i\omega\right)}{\Gamma \left(\frac{1}{2}+\frac{1}{4}\sqrt{1-\frac{4\kappa^2}{M}} \right)\Gamma \left(\frac{1}{2}-\frac{1}{4}\sqrt{1-\frac{4\kappa^2}{M}} \right)}~,  \notag \\
A_{2}  & \approx & C_{1} \frac{\Gamma \left(1-i\omega \right) \Gamma\left(-i\omega \right)}{\Gamma \left(\frac{1}{2}-i\omega-\frac{1}{4}\sqrt{1-\frac{4\kappa^2}{M}} \right)\Gamma \left(\frac{1}{2}-i\omega+\frac{1}{4}\sqrt{1-\frac{4\kappa^2}{M}} \right)}~.
\end{eqnarray}

We will consider the case $4\kappa^2/M<1$. The case $4\kappa^2/M>1$ is similar and the same asymptotic behavior is obtained.

Using the asymptotic expansion of the $\Gamma$ functions for $\left| y \right| \rightarrow \infty$ \cite{AA},

\begin{equation} \label{asymptotic}
\left|\Gamma \left(x+iy \right) \right|=\sqrt{2\pi} \left|y \right| ^{x-1/2} e^{-x-\left| y \right| \pi/2} \left[1+\mathcal{O} \left( \frac{1}{\left| y \right|}\right) \right]~,
\end{equation}

we obtain

\begin{eqnarray}
\left| A_{1}\right| & \approx & \left| C_{1} \right| \frac{2\pi e^{-1-\omega \pi}}{\left| \Gamma \left(\frac{1}{2}+\frac{1}{4}\sqrt{1-\frac{4\kappa^2}{M}} \right) \right| \left| \Gamma \left(\frac{1}{2}-\frac{1}{4}\sqrt{1-\frac{4\kappa^2}{M}} \right) \right|} \propto \frac{1}{e^{\omega \pi}}~, \notag \\
\left| A_{2}\right| & \approx & \left| C_{1} \right| ~,
\end{eqnarray}

therefore, at high frequencies

\begin{eqnarray}
\mathcal{R}  & \propto & \frac{1}{e^{2\omega \pi}} \rightarrow 0~, \notag \\
\mathcal{T}  & \rightarrow & 1~, \notag \\
\sigma & \propto & \frac{1}{\omega} \rightarrow 0~.
\end{eqnarray}

\begin{figure}[h]
\begin{center}
\includegraphics[width=0.7\textwidth]{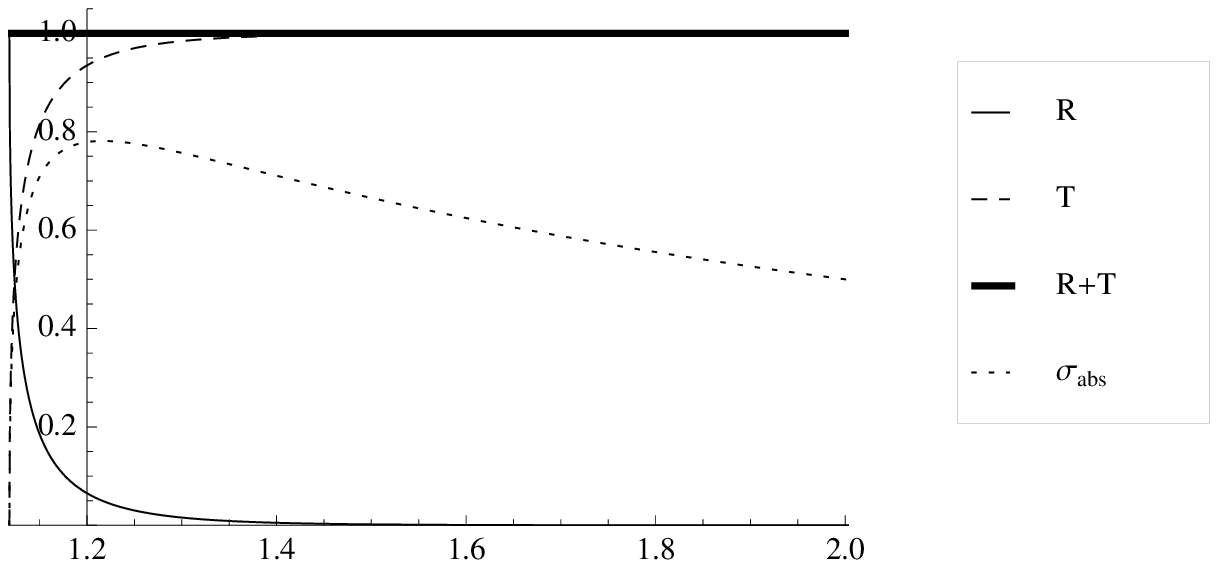}
\includegraphics[width=0.7\textwidth]{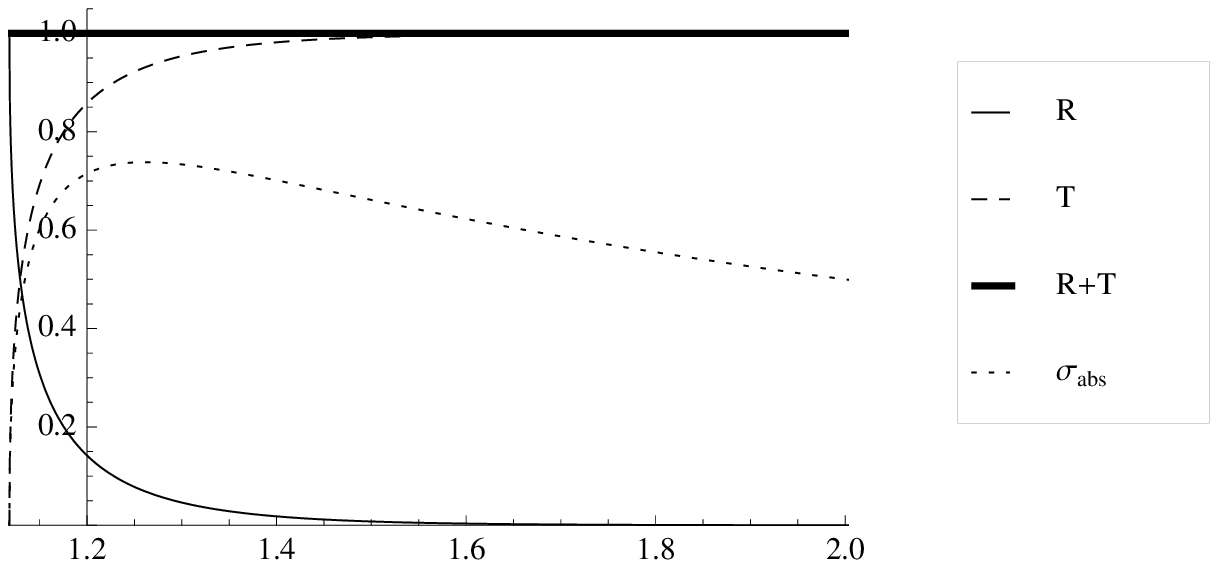}
\includegraphics[width=0.7\textwidth]{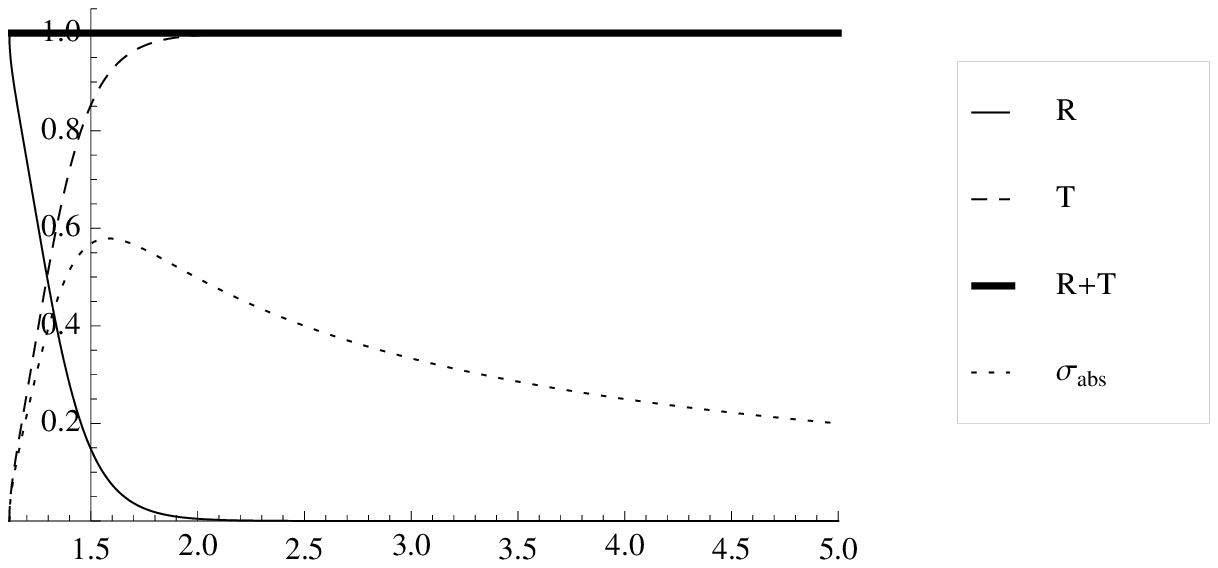}

\end{center}
\caption{The reflection coefficient $\mathcal{R}$ (solid curve), the transmission
coefficient $\mathcal{T}$ (dashed curve), $\mathcal{R}+\mathcal{T}$ (thick curve), and the absorption
cross section $\protect\sigma _{abs}$ (dotted curve) as a function of $%
\protect\omega $, $(1.12\leq \protect\omega )$; for $M=1$ and $m=1$, $\kappa=0$ (upper figure), $\kappa=1$ (middle figure), and $\kappa=2$ (lower figure).} 
\label{figura1}
\end{figure}

Now, we will carry out a numerical analysis of the reflection coefficient~(%
\ref{coef12}), transmission coefficient~(\ref{coef22}), and absorption cross
section~(\ref{absorptioncrosssection2}) of $z=0$ Lifshitz black
holes, for scalar fields. So, we plot the reflection and transmission
coefficients and the absorption cross section in Fig.~(\ref{figura1}) %
, with $M=1$, $m=1$ and $\kappa=0, 1, 2$. We can see, as expected from the above asymptotic analysis, that  the reflection coefficient is one at the low frequency limit, and at the high frequency limit this
coefficient is null. The behavior of the transmission coefficient is
the opposite, with $\mathcal{R}+\mathcal{T}=1$. Also, the absorption cross section is null at the
low and high-frequency limits, but there is a range of frequencies for which
the absorption cross section is not null, and it also has a maximum value.

\begin{figure}[h]
\begin{center}
\includegraphics[width=0.7\textwidth]{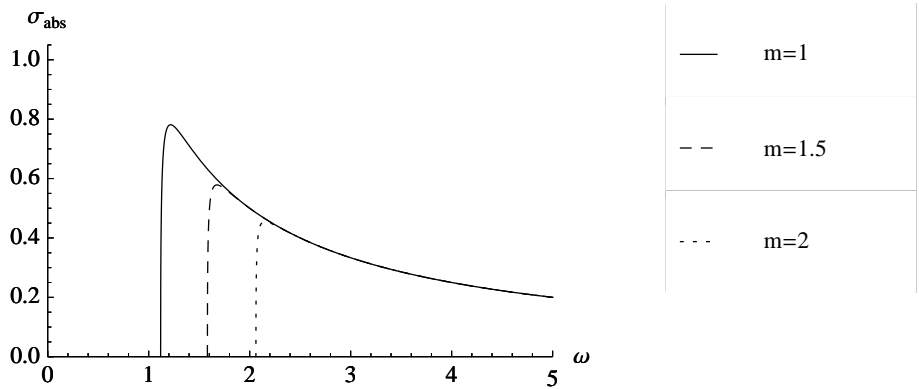}
\end{center}
\caption{The absorption
cross section $\protect\sigma _{abs}$ as a function of $%
\protect\omega $; for $M=1$, $\kappa=0$ and $m=1, 1.5, 2$.}
\label{figura2}
\end{figure}

\begin{figure}[h]
\begin{center}
\includegraphics[width=0.7\textwidth]{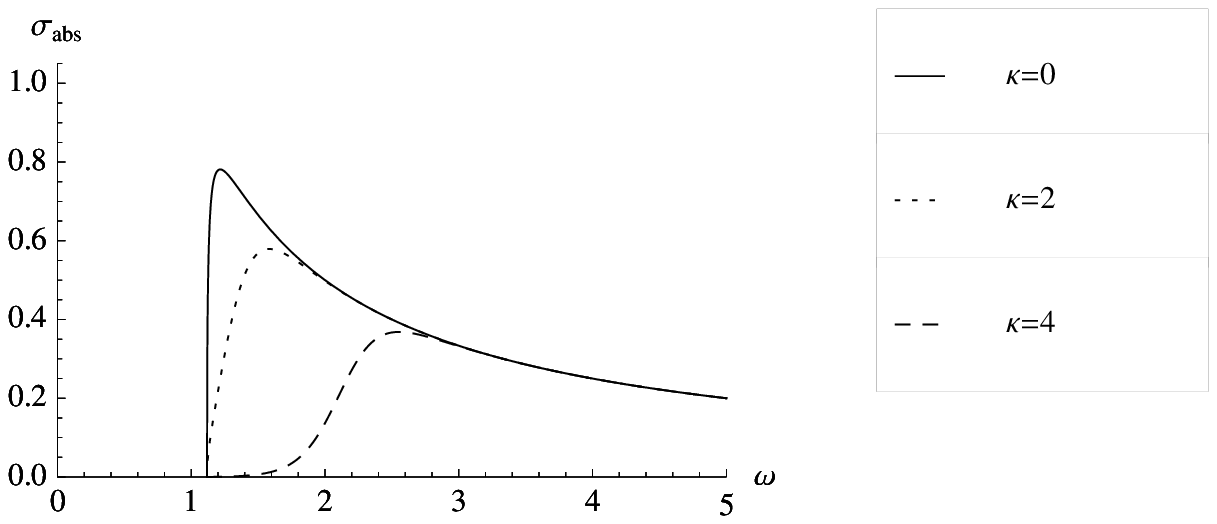}
\end{center}
\caption{The absorption
cross section $\protect\sigma _{abs}$ as a function of $%
\protect\omega $; for $M=1$, $m=1$ and $\kappa=0, 2, 4$.}
\label{figura3}
\end{figure}

The QNFs are defined as the poles of the transmission coefficient, which is equivalent to impose that only outgoing waves exist at the asymptotic infinity. These poles
are given by $A_{2}=0$, and this occurs when $c-a+n=0$ or $c-b+n=0$, with $%
n=0,1,2$ ... . Therefore, the quasinormal frequencies are given by 
\begin{equation}
\omega= -i\frac{-2m^{2}\left( 1+2n\right) +1+10n+24n^{2}+16n^{3}+2%
\frac{\kappa ^{2}}{M}+4n\frac{\kappa ^{2}}{M}\pm\sqrt{1-4\frac{\kappa
^{2}}{M}}\left( 1+m^{2}+4n+4n^{2}+\frac{\kappa ^{2}}{M}\right) }{%
3+16n+16n^{2}+4\frac{\kappa ^{2}}{M}}~.
\end{equation}

We observe that for $1-4\frac{\kappa^2}{M} > 0$ the QNFs are purely imaginary and for $1-4\frac{\kappa^2}{M} < 0$ they acquire a real part. Also, we observe that for some
values of the scalar field mass $m$ the quasinormal frequencies can have a positive imaginary part, therefore we conclude that this black hole is not stable under scalar field perturbations minimally coupled to gravity.

\subsection{Scalar field conformally coupled to gravity}

 In this section we consider a scalar field non-minimally coupled to curvature, propagating in the background of the Lifshitz black hole (\ref{solution}). In particular, we study the case of a scalar field conformally coupled. The Klein-Gordon equation is given by

\begin{equation}
\frac{1}{\sqrt{-g}}\partial _{\mu }\left( \sqrt{-g}g^{\mu \nu }\partial
_{\nu }\right) \phi -\chi R \phi=m^{2}\phi ~,
\end{equation}

where $\chi$ is the non-mininal coupling parameter, and $R=-2\left( 1-\frac{M}{r^2}\right)$ is the Ricci scalar.

In this case, by means of the ansatz $\phi=e^{-i\omega t}e^{i\kappa \theta}R(r)$, we obtain the following radial equation

\begin{equation}
r^{2}\left( r^{2}-M\right) \frac{d^{2}R\left(
r\right)}{dr^{2}}+2r^{3}\frac{dR\left(
r\right)}{dr}+\left( 
\frac{\omega ^{2}r^{4}}{r^{2}-M}-\left( \kappa ^{2}+2\chi M \right) - \left( m^{2}-2\chi \right)  r^{2}\right) R\left(
r\right) =0~.
\end{equation}%
We can see that, due to the simplicity of the Ricci scalar in this background, it is possible to obtain this equation directly from (\ref{first}) by means of the following transformations
\begin{eqnarray} \label{cambio}
\kappa^{2} & \rightarrow & \kappa^2+2\chi M~, \notag \\
m^2  & \rightarrow & m^2-2\chi~.
\end{eqnarray}

Therefore, the reflection and transmission coefficients, the absorption cross section and the quasinormal modes for the non-minimal case can be obtained easily from the equations (\ref{coef12}), (\ref{coef22}) and (\ref{absorptioncrosssection2}) of the minimal case by using transformations (\ref{cambio}).

For a conformally coupled scalar field, $m=0$ and $\chi=1/8$, one can find that the qualitative behavior of the reflection and transmission coefficients and the absortion cross section is similar to the obtained in section $A$ for a minimally coupled scalar field, and for the quasinormal frequencies we obtain the following expression

\begin{equation}
\omega= \pm\frac{\kappa}{2\sqrt{M}} -i\frac{1+6n+12n^{2}+8n^{3}+%
\frac{\kappa ^{2}}{M}+2n\frac{\kappa ^{2}}{M}}{%
2+8n+8n^{2}+2\frac{\kappa ^{2}}{M}}~,
\end{equation}

which have a negative imaginary part, therefore, the black hole is stable under conformally coupled scalar field perturbations.

Also, we note that the effective potential  goes to zero at the asymptotic infinity.

\section{Final Remarks}

In this work we studied scalar field perturbations of an asymptotically Lifshitz black hole in three-dimensional conformal gravity with dynamical exponent $z=0$,
where in this case the anisotropic scale invariance corresponds to a space-like scale
invariance with no transformation of time, and we calculated the reflection and transmission coefficients, the absorption cross section and the quasinormal modes. The results obtained show that the absorption cross section vanishes at the low frequency limit as well as at the high frequency limit. This means that a wave emitted from the horizon, with low or high frequency, does not reach spatial infinity and therefore is totally reflected, because the fraction of particles  penetrating the potential barrier vanishes. We have also shown in Fig. (\ref{figura1}) that there is a range of frequencies where the absorption cross section is not null. On the other hand, the reflection coefficient is one at the low frequency limit and null at high frequencies; the behavior of the transmission coefficient is the opposite, where $\mathcal{R}+\mathcal{T}=1$. Furthermore, we have shown that the absorption cross section decreases for higher values of angular momentum, and decreases when the mass $m$ of the scalar field increases; however, for high frequencies the difference is negligible, see Figs. (\ref{figura2}, \ref{figura3}).

Furthermore, we calculated analytically the QNFs of scalar
perturbations, which coincide with the poles of the transmission coefficient, and we
found two sets of quasinormal frequencies. However, some of these can have a positive imaginary part, depending of the value of $m$, and therefore the black hole is not stable under scalar field perturbations for a scalar field minimally coupled to curvature. If the scalar field is conformally coupled to the curvature, we found that the imaginary part of the quasinormal frequencies is negative, which guarantees the linear stability of the black hole in this case.

Interesting aplications we hope to address in the near future are the holographic implications of the results of this paper.

\section*{Acknowledgments}

This work was funded by the Comisi{\'o}n Nacional de Investigaci{\'o}n Cient{%
\'i}fica y Tecnol{\'o}gica through FONDECYT Grant 11121148 (YV, MC) and also
partially funded by Direcci{\'o}n de investigaci{\'o}n, Universidad de La
Frontera (MC).

\appendix


\begin{thebibliography}{99}

\bibitem{Banados:1992wn}  M.~Banados, C.~Teitelboim and J.~Zanelli,  
Phys.\ Rev.\ Lett.\ \textbf{69} (1992) 1849  [hep-th/9204099].  


\bibitem{Deser:1981wh}  S.~Deser, R.~Jackiw and S.~Templeton,  
Annals Phys.\ \textbf{140} (1982) 372  [Erratum-ibid.\ \textbf{185} (1988)
406]  [Annals Phys.\ \textbf{185} (1988) 406]  [Annals Phys.\ \textbf{281}
(2000) 409].  


\bibitem{Deser:1982vy}  S.~Deser, R.~Jackiw and S.~Templeton,  
Phys.\ Rev.\ Lett.\ \textbf{48} (1982) 975.  


\bibitem{Bergshoeff:2009hq}  E.~A.~Bergshoeff, O.~Hohm and P.~K.~Townsend,  
Phys.\ Rev.\ Lett.\ \textbf{102} (2009) 201301  [arXiv:0901.1766 [hep-th]].  


\bibitem{Bergshoeff:2009tb}  E.~A.~Bergshoeff, O.~Hohm and P.~K.~Townsend,  
Annals Phys.\ \textbf{325} (2010) 1118  [arXiv:0911.3061 [hep-th]].  


\bibitem{Andringa:2009yc}  R.~Andringa, E.~A.~Bergshoeff, M.~de Roo,
O.~Hohm, E.~Sezgin and P.~K.~Townsend,  
Class.\ Quant.\ Grav.\ \textbf{27} (2010) 025010  [arXiv:0907.4658
[hep-th]].  


\bibitem{Clement:2009gq}  G.~Clement,  
Class.\ Quant.\ Grav.\ \textbf{26} (2009) 105015  [arXiv:0902.4634
[hep-th]].  


\bibitem{AyonBeato:2009yq}  E.~Ayon-Beato, G.~Giribet and M.~Hassaine,  
JHEP \textbf{0905} (2009) 029  [arXiv:0904.0668 [hep-th]].  


\bibitem{Clement:2009ka}  G.~Clement,  
Class.\ Quant.\ Grav.\ \textbf{26} (2009) 165002  [arXiv:0905.0553
[hep-th]].  

\bibitem{Oliva:2009ip}  J.~Oliva, D.~Tempo and R.~Troncoso,  
JHEP \textbf{0907}, 011 (2009)  [arXiv:0905.1545 [hep-th]].

\bibitem{Correa:2014ika}
  F.~Correa, M.~Hassaine and J.~Oliva,
  arXiv:1403.6479 [hep-th].


\bibitem{AyonBeato:2009nh}  E.~Ayon-Beato, A.~Garbarz, G.~Giribet and
M.~Hassaine,  
Phys.\ Rev.\ D \textbf{80} (2009) 104029  [arXiv:0909.1347 [hep-th]].  



\bibitem{Nakasone:2009bn}  M.~Nakasone and I.~Oda,  
Prog.\ Theor.\ Phys.\ \textbf{121} (2009) 1389  [arXiv:0902.3531 [hep-th]].  



\bibitem{Bergshoeff:2009aq}  E.~A.~Bergshoeff, O.~Hohm and P.~K.~Townsend,  
Phys.\ Rev.\ D \textbf{79} (2009) 124042  [arXiv:0905.1259 [hep-th]].  


\bibitem{Oda:2009ys}  I.~Oda,  
JHEP \textbf{0905} (2009) 064  [arXiv:0904.2833 [hep-th]].  



\bibitem{Deser:2009hb}  S.~Deser,  
Phys.\ Rev.\ Lett.\ \textbf{103} (2009) 101302  [arXiv:0904.4473 [hep-th]].  

\bibitem{Oliva:2009hz} 
  J.~Oliva, D.~Tempo and R.~Troncoso,
  Int.\ J.\ Mod.\ Phys.\ A {\bf 24}, 1588 (2009)
  [arXiv:0905.1510 [hep-th]].


\bibitem{Maldacena:1997re} J.~M.~Maldacena, 
Adv.\ Theor.\ Math.\ Phys.\ \textbf{2}, 231 (1998) [hep-th/9711200]. 

\bibitem{Kachru:2008yh} S.~Kachru, X.~Liu and M.~Mulligan, 
Phys.\ Rev.\ D \textbf{78}, 106005 (2008) [arXiv:0808.1725 [hep-th]]. 




\bibitem{Hartnoll:2009ns} S.~A.~Hartnoll, J.~Polchinski, E.~Silverstein and
D.~Tong, 
JHEP \textbf{1004}, 120 (2010) [arXiv:0912.1061 [hep-th]]. 




\bibitem{Cai:2009ac} R.~-G.~Cai, Y.~Liu and Y.~-W.~Sun, 
JHEP \textbf{0910} (2009) 080 [arXiv:0909.2807 [hep-th]]. 


\bibitem{AyonBeato:2010tm} E.~Ayon-Beato, A.~Garbarz, G.~Giribet and
M.~Hassaine, 
JHEP \textbf{1004} (2010) 030 [arXiv:1001.2361 [hep-th]]. 


\bibitem{Dehghani:2010kd} M.~H.~Dehghani and R.~B.~Mann, 
JHEP \textbf{1007} (2010) 019 [arXiv:1004.4397 [hep-th]]. 


\bibitem{Mann:2009yx} R.~B.~Mann, 
JHEP \textbf{0906} (2009) 075 [arXiv:0905.1136 [hep-th]]. 


\bibitem{Balasubramanian:2009rx}  K.~Balasubramanian and J.~McGreevy,  
Phys.\ Rev.\ D \textbf{80} (2009) 104039  [arXiv:0909.0263 [hep-th]].  


\bibitem{Bertoldi:2009vn} G.~Bertoldi, B.~A.~Burrington and A.~Peet, 
Phys.\ Rev.\ D \textbf{80} (2009) 126003 [arXiv:0905.3183 [hep-th]]. 


\bibitem{Bravo-Gaete:2013dca}  M.~Bravo-Gaete and M.~Hassaine,  
arXiv:1312.7736 [hep-th].  


\bibitem{Alvarez:2014pra}  A.~Alvarez, E.~Ayon-Beato, Hernan A.~Gonzalez and
M.~Hassaine,  
arXiv:1403.5985 [gr-qc].  


\bibitem{Stelle:1976gc}  K.~S.~Stelle,  
Phys.\ Rev.\ D \textbf{16} (1977) 953.  


\bibitem{Stelle:1977ry}  K.~S.~Stelle,  
Gen.\ Rel.\ Grav.\ \textbf{9} (1978) 353.  

\bibitem{Dirac} P. A. M. Dirac, �Lectures on Quantum Mechanics�, Belfer
Graduate School of Science,Yeshiva University Press, New York (1964).


\bibitem{Pais:1950za}  A.~Pais and G.~E.~Uhlenbeck,  
Phys.\ Rev.\ \textbf{79} (1950) 145.  


\bibitem{Mannheim:2006rd}  P.~D.~Mannheim,  
Found.\ Phys.\ \textbf{37} (2007) 532  [hep-th/0608154].  

\bibitem{Lu:2012xu} 
  H.~Lu, Y.~Pang, C.~N.~Pope and J.~F.~Vazquez-Poritz,
  Phys.\ Rev.\ D {\bf 86}, 044011 (2012)
  [arXiv:1204.1062 [hep-th]].

\bibitem{Guralnik:2003we} 
  G.~Guralnik, A.~Iorio, R.~Jackiw and S.~Y.~Pi,
  Annals Phys.\  {\bf 308}, 222 (2003)
  [hep-th/0305117].

\bibitem{Grumiller:2003ad} 
  D.~Grumiller and W.~Kummer,
  Annals Phys.\  {\bf 308}, 211 (2003)
  [hep-th/0306036].

\bibitem{Maldacena:1996ix}
  J.~M.~Maldacena and A.~Strominger,
  Phys.\ Rev.\ D {\bf 55} (1997) 861
  [hep-th/9609026].


\bibitem{Harmark:2007jy}
T.~Harmark, J.~Natario and R.~Schiappa,
Adv.\ Theor.\ Math.\ Phys.\  {\bf 14}, 727 (2010)
[arXiv:0708.0017 [hep-th]].


\bibitem{Moderski:2008nq}
  R.~Moderski and M.~Rogatko,
  Phys.\ Rev.\ D {\bf 77} (2008) 124007
  [arXiv:0805.0665 [hep-th]].
  
\bibitem{Gibbons:2008gg}
  G.~W.~Gibbons, M.~Rogatko and A.~Szyplowska,
  Phys.\ Rev.\ D {\bf 77} (2008) 064024
  [arXiv:0802.3259 [hep-th]].
  

\bibitem{Regge:1957td} T.~Regge and J.~A.~Wheeler, 
Phys.\ Rev.\ \textbf{108}, 1063 (1957).


\bibitem{Zerilli:1971wd} F.~J.~Zerilli, 
Phys.\ Rev.\ D \textbf{2}, 2141 (1970).


\bibitem{Zerilli:1970se} F.~J.~Zerilli, 
Phys.\ Rev.\ Lett.\ \textbf{24}, 737 (1970). 


\bibitem{Kokkotas:1999bd} K.~D.~Kokkotas and B.~G.~Schmidt, 
Living Rev.\ Rel.\ \textbf{2}, 2 (1999) [gr-qc/9909058]. 


\bibitem{Nollert:1999ji} H.~-P.~Nollert, 
Class.\ Quant.\ Grav.\ \textbf{16}, R159 (1999). 



\bibitem{Konoplya:2011qq} R.~A.~Konoplya and A.~Zhidenko, 
Rev.\ Mod.\ Phys.\ \textbf{83}, 793 (2011) [arXiv:1102.4014 [gr-qc]].


\bibitem{Birmingham:2001pj} D.~Birmingham, I.~Sachs and S.~N.~Solodukhin, 
Phys.\ Rev.\ Lett.\ \textbf{88}, 151301 (2002) [hep-th/0112055].

\bibitem{Corda:2012tz} 
  C.~Corda,
  Int.\ J.\ Mod.\ Phys.\ D {\bf 21}, 1242023 (2012)
  [arXiv:1205.5251 [gr-qc]].

\bibitem{Corda:2012dw} 
  C.~Corda,
  Eur.\ Phys.\ J.\ C {\bf 73}, 2665 (2013)
  [arXiv:1210.7747 [gr-qc]].

\bibitem{Chan:1996yk} 
  J.~S.~F.~Chan and R.~B.~Mann,
  Phys.\ Rev.\ D {\bf 55}, 7546 (1997)
  [gr-qc/9612026].

\bibitem{Cardoso:2001hn} 
  V.~Cardoso and J.~P.~S.~Lemos,
  Phys.\ Rev.\ D {\bf 63}, 124015 (2001)
  [gr-qc/0101052].


\bibitem{Kwon:2011ey}  Y.~Kwon, S.~Nam, J.~-D.~Park and S.~-H.~Yi,  
Class.\ Quant.\ Grav.\ \textbf{28} (2011) 145006  [arXiv:1102.0138
[hep-th]].  


\bibitem{Gonzalez:2014voa} 
  P.~A.~Gonzalez and Y.~Vasquez,
  arXiv:1404.5371 [gr-qc].



\bibitem{CuadrosMelgar:2011up} B.~Cuadros-Melgar, J.~de Oliveira and
C.~E.~Pellicer, 
Phys.\ Rev.\ D \textbf{85}, 024014 (2012) [arXiv:1110.4856 [hep-th]]. 


\bibitem{Gonzalez:2012de} P.~A.~Gonzalez, J.~Saavedra and Y.~Vasquez, 
Int.\ J.\ Mod.\ Phys.\ D \textbf{21}, 1250054 (2012) [arXiv:1201.4521
[gr-qc]]. 

\bibitem{Myung:2012cb} Y.~S.~Myung and T.~Moon, 
Phys.\ Rev.\ D \textbf{86} (2012) 024006 [arXiv:1204.2116 [hep-th]]. 


\bibitem{Gonzalez:2012xc} P.~A.~Gonzalez, F.~Moncada and Y.~Vasquez, 
Eur.\ Phys.\ J.\ C \textbf{72}, 2255 (2012) [arXiv:1205.0582 [gr-qc]]. 


\bibitem{Becar:2012bj} R.~Becar, P.~A.~Gonzalez and Y.~Vasquez, 
Int.\ J.\ Mod.\ Phys.\ D \textbf{22}, 1350007 (2013) [arXiv:1210.7561
[gr-qc]]. 


\bibitem{Giacomini:2012hg} A.~Giacomini, G.~Giribet, M.~Leston, J.~Oliva and
S.~Ray, 
Phys.\ Rev.\ D \textbf{85} (2012) 124001 [arXiv:1203.0582 [hep-th]]. 

\bibitem{Catalan:2014ama} 
  M.~Catalan, E.~Cisternas, P.~A.~Gonzalez and Y.~Vasquez,
  arXiv:1404.3172 [gr-qc].


\bibitem{Catalan:2013eza}  M.~Catalan, E.~Cisternas, P.~A.~Gonzalez and
Y.~Vasquez,  
arXiv:1312.6451 [gr-qc].  

\bibitem{Lopez-Ortega:2014daa} 
  A.~Lopez-Ortega,
  arXiv:1407.0966 [gr-qc].

\bibitem{Lopez-Ortega:2014oha} 
  A.~L\'opez-Ortega,
  Gen.\ Rel.\ Grav.\  {\bf 46}, 1756 (2014)
  [arXiv:1406.0126 [gr-qc]].

\bibitem{Moon:2012dy} 
  T.~Moon and Y.~S.~Myung,
  Eur.\ Phys.\ J.\ C {\bf 72}, 2186 (2012)
  [arXiv:1205.2317 [hep-th]].

\bibitem{Lepe:2012zf}  S.~Lepe, J.~Lorca, F.~Pena and Y.~Vasquez,  
Phys.\ Rev.\ D \textbf{86} (2012) 066008  [arXiv:1205.4460 [hep-th]].  


\bibitem{Olivares:2013zta}  M.~Olivares, Y.~Vasquez, J.~R.~Villanueva and
F.~Moncada,  
arXiv:1306.5285 [gr-qc].  


\bibitem{Olivares:2013uha}  M.~Olivares, German Rojas, Y.~Vasquez and
J.~R.~Villanueva,  
Astrophys.\ Space Sci.\ \textbf{347} (2013) 83  [arXiv:1304.4297 [gr-qc]].  


\bibitem{Villanueva:2013gra}  J.~R.~Villanueva and Y.~Vasquez,  
Eur.\ Phys.\ J.\ C \textbf{73} (2013) 2587  [arXiv:1309.4417 [gr-qc]].  


\bibitem{Alishahiha:2012nm} M.~Alishahiha, M.~R.~Mohammadi Mozaffar and
A.~Mollabashi, 
Phys.\ Rev.\ D \textbf{86} (2012) 026002 [arXiv:1201.1764 [hep-th]]. 

\bibitem{Lu:2013hx} 
  H.~Lü, Y.~Pang and C.~N.~Pope,
  Phys.\ Rev.\ D {\bf 87}, no. 10, 104013 (2013)
  [arXiv:1301.7083 [hep-th]].

\bibitem{M. Abramowitz} M. Abramowitz and A. Stegun, Handbook of
Mathematical functions, (Dover publications, New York, 1970).

\bibitem{AA}
A. Kilbas and S. Megumi. A Remark on Asymptotics of the Gamma Function at Infinity (Study on Applications for Fractional Calculus Operators in Univalent Function Theory). http://hdl.handle.net/2433/25303

\end{thebibliography}
\end{document}